\documentclass[12pt]{article}
\usepackage[]{graphicx}
\newcommand{\MET}{\mbox{$E_T\hspace{-0.237in}\not\hspace{0.18in}$}}
\begin{document}\setlength{\unitlength}{1mm}

\begin{titlepage}

\begin{flushright}
{RUNHETC-2011-05}\\
\end{flushright}

\vskip2cm
\begin{center}
{\large {\bf LHC Charge Asymmetry as Constraint}}
\\
{\ }
\\
{\large {\bf on Models for the Tevatron Top Anomaly}}
\vskip1.0cm {Nathaniel Craig,$^{a,b}$ Can Kilic,$^{b}$ and Matthew~J.~Strassler$^{b}$}
\vskip0.5cm
{\it $^{a}$ School of Natural Sciences, Institute for Advanced Study,\\ Einstein Drive, Princeton, NJ 08540, USA}\\
\vskip0.5cm
{\it $^{b}$ New High Energy Theory Center,\\ Department of Physics and Astronomy,  Rutgers University, \\ 136 Frelinghuysen Rd, Piscataway, NJ 08854}

\end{center}

\vspace{0.5cm}
\begin{abstract}
The forward-backward asymmetry $A_{FB}^{t\bar t}$ in top quark
production at the Tevatron has been observed to be anomalously large
by both CDF and D0.  It has been suggested that a model with a $W'$
coupling to $td$ and $ub$ might explain this anomaly, and other anomalies in
$B$ mesons.  Single-top-quark production in this model is large, and
arguably in conflict with Tevatron measurements.  However the model
might still be viable if $A_{FB}^{t\bar t}$ is somewhat smaller than
its current measured central value.  We show that even with smaller
couplings, the model can be discovered (or strongly excluded) at the
LHC using the 2010 data sets.  We find that a suitable
charge-asymmetry measurement is a powerful tool that can be used to
constrain this and other sources of anomalous single-top production,
and perhaps other new high-energy charge-asymmetric processes.
\end{abstract}

\end{titlepage}

The forward-backward asymmetry in top quark pair-production has been
measured by CDF and D0 to be anomalously large \cite{CDFafb, D0afb,
  CDFnew}.  It seems difficult to explain the size and nature of the
asymmetry using Standard Model physics, Monte Carlo subtleties, or
experimental difficulties.  Various models of new physics have been proposed, in which a
new particle contributes to $t\bar t$ production.
Most of these have problems with fitting other data.  For instance,
the insertion of a $Z'$ that couples to $u$ and $t$ quarks, allowing
for $u\bar u\to t\bar t$, creates a large rate for $u u \to t t$.
This is especially true at the LHC, where both the $u$ quarks are
valence quarks.  The corresponding signal of two same-sign leptons has
hardly any Standard Model background and is easily excluded for a $Z'$
of the required mass  \cite{Hitoshi, Wagner}.

One model recently proposed by Shelton and Zurek \cite{SZ} involves a
similar structure, but with a $W'$ exchange.  The $W'$ considered is
called ``maximally flavor violating'' --- one might rather say
``maximally generationally violating'' -- in that it couples
right-handed quarks $u$ to $b$, and also $t$ to $d$, with all other
couplings strongly suppressed to avoid new sources of flavor-changing
neutral currents.\footnote{Various $W'$ proposals for explaining
the forward-backward asymmetry involving a $W'td$ coupling but
 no $W'ub$ coupling were considered previously in
\cite{Cheung:2009ch, Cheung, Barger:2010mw, Wagner}.}  The
$W'$ is imagined to be of order 600 GeV, with
the $Z'$ considerably heavier to be consistent with precision
electroweak measurements.  Note the $Z'$ has little or no
flavor-changing couplings and does not contribute to low-background
observables at the Tevatron or LHC.  The $W'td$ coupling is necessary
to explain the $t\bar t$ asymmetry.  But the authors of \cite{SZ}
suggest further that a $W'ub$ coupling could explain (at least)
several anomalies in the $B$ meson system: the D0 measurement of the
like-sign dimuon charge asymmetry in semileptonic $b$ decays
\cite{D0dimuon}; the deviation of $B_s - \overline{B}_s$ mixing from
Standard Model predictions in measurements of $\Delta \Gamma_s$ and
$S_{\psi \phi}$ by both D0 \cite{D0Bs} and CDF \cite{CDFBs}; and
indications of new CP violation in the $B_d$ system in $B_d \to \psi
K_s$ \cite{Lunghi, Ligeti} and $B_d \to (\phi,\eta',\pi,\rho,\omega)
K_s$ \cite{Barger}.

However, in the presence of a $W'ub$ coupling with the same strength
as a $W'td$ coupling, the same logic that limits a $Z'ut$ coupling
potentially applies to a $W'$.  A new source of single-top quark
production, through the processes $ub\to td$ and $ud\to tb$ (and their
conjugates), becomes possible via $W'$ exchange.  The $t$-channel $W'$
exchange process, $ud\to tb$, can proceed from a color-octet initial
state and can be large at the Tevatron, even for a heavy $W'$.  At
the LHC, meanwhile, this process is enormous, due to the fact that
both quarks in the initial state are from valence distributions.
Meanwhile the huge gluon flux at small $x$ and the accessibility of
the $W'$ resonance means that the color-singlet $s$-channel process
$ub\to td$ is also quite large at the LHC.

Despite this large cross-section, the final state contains only one
lepton, and is not as distinctive as the same-sign dileptons arising
in the $uu\to tt$ case.  There are therefore large backgrounds from
$W$-plus-jets and from $t\bar t$.

What makes this signal extraordinary --- also true of the
$uu\to tt$ signal in the $Z'ut$ model --- is its charge asymmetry at the LHC.  In comparison
to single-top production in the Standard Model, which already has
substantial asymmetries (a forward-backward asymmetry at the Tevatron
and a roughly 2:1 charge asymmetry at the 7 TeV LHC), single top
production in the $W'$ model has an LHC charge asymmetry of
order 16:1.  This can be put to use, applying a variant of
the simple but powerful method that was suggested by Bowen \cite{Bowen}
(following \cite{BES}) for measuring single top in the Standard Model
(SM) at the LHC.

The use of charge asymmetries at $pp$ colliders has been discussed
actively in the past.  Examples have appeared in the literature on
supersymmetry, which can give observable asymmetric signals; see for
example \cite{asymsusy}.  The use of charge asymmetries in
SM single top searches was suggested in \cite{Slabospitsky} at the UNK collider,
prior to the independent work of \cite{Bowen} for the LHC.  The need to apply
charge asymmetries systematically for new physics searches has been
argued for by one of us \cite{Pheno2005}, and independently by Stirling and Kom
\cite{Stirling}, who have performed a serious investigation of SM
backgrounds.  The current discussion of new models to explain the
forward-backward top asymmetry at the Tevatron now provides us a first
opportunity to put these variables in play at the LHC.

\

The rate for single-top production in the $W'$ model depends
on the $W'$ mass $M_{W'}$ and its coupling $g_R$ to $td$ and $ub$.
In \cite{SZ} the preferred $W'$ mass was about 600 GeV and the
coupling $g_R$ was preferred in the range 1.5 to 2, following \cite{Cheung}.
We will take the coupling $g_R=1.5$ and $M_{W'}=600$ GeV as the ``fiducial
values'' for the parameters, and call this the ``fiducial point''
in parameter space.

Before exploring the signal at the LHC,
let us first consider it at the Tevatron.
At  the fiducial point,
we find that single-top production at the Tevatron is
increased, relative to the Standard Model, by a factor of 2, most of
it in the $t$ channel.  (Note there is no interference with standard
model single-top production, which has a final state antiquark.)
Here
we are taking the leading-order new-physics result and comparing it to
the next-to-leading-order (NLO) Standard Model single-top
cross-section; the $K$ factor for the new physics is likely above 1,
so we are probably conservative by taking it to be $\sim 1$.  If $g_R$ were
2, the rate for single-top production would grow to 5 times the SM
prediction.  Uncertainties on the measured cross-section at CDF and D0 are
relatively small, of order 25\% of the Standard Model
cross-section \cite{GroupST}.  We therefore believe that $g_R\sim 2$ is already
strongly excluded, and $1.5$ is considerably disfavored.

Yet the situation is difficult to interpret just with cross-sections,
because the single top signal at the Tevatron is extracted using a
complex multivariate analysis from a very large background, assuming
the shape of the signal is that of the SM.  The addition of the new
single-top signal from $ud\to tb$ and its conjugate to the SM
processes will change that shape, so the analysis must be repeated
by those who performed it originally.

That said, it seems likely that the model at its fiducial point would
already have revealed itself through a single-top excess at the
Tevatron.  But the fiducial values of the parameters were chosen in \cite{SZ, Cheung} to
fit the central value of the CDF measurement of $A_{FB}^{t\bar t}$,
which is very large, but has a large statistical error bar.  For the
usual reasons, it may be expected that the true value of $A_{FB}^{t\bar
  t}$ is lower than the current central value.  The $W'$ model might
then survive, and still explain the $A_{FB}^{t\bar t}$ data, with a
slightly larger mass and/or smaller coupling constant.  Moreover,
since the effect on $A_{FB}^{t\bar t}$ is through interference, while
the single-top measurement is the square of a non-interfering
amplitude, a reduction in the asymmetry by a factor $z$ is accompanied
by a reduction in $t$-channel single-top production by roughly a factor of $z^2$.

Furthermore, as a sociological statement, one might note that
single-top production was predicted with precision in the SM well in
advance of its observation at the Tevatron, and thus there is no truly unbiased
measurement of this process.  The measurement is complicated, and hard
to check by eye in a single plot.  We might wish to remain a bit
cautious until the results are confirmed by an entirely different
technique.

Therefore, while we would view the $W'$ model as disfavored somewhat,
it does not seem to us to be obviously excluded.  A much more detailed
Tevatron study would be needed, and arguments might still ensue as to
the limits obtained.

\

However, at the LHC it seems possible to discover or exclude the model
more cleanly, using only the existing 2010 data sets of $\sim 35$
inverse pb per experiment.  We will argue below that the charge
asymmetry in a sample consisting of a single lepton, a small amount of
missing transverse momentum ($\MET$, or MET), and at least two jets is
already sensitive to signals of this type.  Application of simple
kinematic cuts and/or heavy-flavor tagging permits an excess charge
asymmetry to be observed even for a signal much smaller than arises in
the fiducial case.  This in turn means that the coupling and mass of
the $W'$ can be strongly constrained by this measurement.

For the fiducial point, we find that the LO production cross-section
$\sigma_{t}^{(0)}$ for single top quarks from $W'$ exchange is 220 pb.
About two-thirds of the cross-section comes from $t$-channel $W'$
exchange, through $ud\to tb$ and its conjugate, and has a 20:1 charge
asymmetry.  The remainder goes through $ub\to td$, through the $W'$ in
the $s$-channel.\footnote{The $W'$ resonance has a width of order
  100 GeV, and may even be wider if the $W'$ has other decay modes not
  included in the minimal model. The resonance might be
  reconstructable if the width is small enough, but since the width is
  model-dependent, we will not rely upon it below.  Clearly, if a
  signal is observed, an attempt should be made to search for the
  resonance in $t$-plus-jet.}  This channel has a charge asymmetry of
order 10:1.  There will be considerable corrections to these LO
results, but we do not believe there will be significant reductions.
There is also an interesting $tW'$ process, but it is too small
to affect our discussion.

We are going to show that even a fraction of these LO cross-sections
can easily be observed relative to NLO-rescaled backgrounds.
Since we do not know the NLO correction to the LO estimate, and the
parameters need not be at the fiducial point, we define for
convenience $F_S\equiv\sigma_{t}^{true}/ \sigma_t^{(0)}$ to be the
appropriate normalization constant.  For the most part we do not
expect enormous differences in shapes as parameters vary or due to NLO
corrections; in any case these could be computed in the future.
Initial state
radiation (ISR) can have an effect on some distributions, and we will
account for that as appropriate.
The largest shape variation will occur if $M_{W'}$ is much above
600 GeV; the $s$-channel process, which is subdominant anyway, will
be reduced the most, though this will be somewhat compensated by its
higher-energy kinematic distribution.

Even without using the charge asymmetry, there is good reason to think
that public results from the LHC already exclude the $W'$ model at the
fiducial point.  Distributions of the total numbers of
events with a lepton, MET and three or more jets, versus an effective
mass variable $m_{{\rm eff}}$, have been shown in a recent supersymmetry
search by ATLAS \cite{ATLASSUSY}.\footnote{We thank J.~Ruderman,
  D.~Shih and N.~Toro for suggesting this study might be relevant for
  us.}  The signal region of this supersymmetry search requires large
MET and large transverse mass.  Our signal has a tail out to large
MET, but this comes from a $W$ decay, so it has low transverse mass,
and relatively little will appear in the ATLAS signal region.
However, control samples for this search, with
low MET and low transverse mass, and with either zero or $\geq 1$ $b$ tags,
have been shown \cite{ATLASTalk}.  These have an event selection that
would be somewhat sensitive to this signal.

The signal is so large, and extends to such large values of
$H_T^{\ell\nu jj}$, that it seems at first obvious that $F_S=1$ is
already excluded by the paucity of events at high $H_T$ in the control
regions of the ATLAS search.  More study reveals that the exclusion is
probable but not overwhelming.  The restriction to a low range of MET
($30<\MET<80$ GeV) eliminates of order half our signal, and also pulls
down the $H_T^{\ell\nu j j}$ distribution, reducing the tail at high
values.  The requirement of a third jet removes quite a bit of signal
as well.  A rough estimate suggests that at $F_S=1$ the new single-top
signal would produce
about 10 events above $m_{{\rm eff}}=800$ GeV in the zero-tag control
sample (called the ``$W$ region'').  But the sample shows
no events.  Still, we remain cautious, because extracting a quantitative
limit would require more details of how the control samples were
obtained and normalized, and more information about relevant
efficiencies.  In any case, it does seems likely that $F_S=1$ is
excluded, as at the Tevatron, but $F_S\sim 0.25$ may well not yet be
excluded.  As we have noted, this and even lower values are still
potentially interesting for the $A_{FB}^{t\bar t}$ anomaly.

We should note that our signal might show up more strikingly in the ATLAS control sample with
high MET and low transverse mass.  Unfortunately the plot for this control region was not shown in public.

It is our view that the use of a charge asymmetry, considered as a
function of a variable such as $m_{{\rm eff}}$, with no upper
restriction on the MET, and with no requirement of a third jet, would
be efficient for signal and allow for a much more powerful and
convincing exclusion of the model even if $F_S=0.25$.  In particular,
any excess at high $m_{{\rm eff}}$, if this or any similar model is
correct, should be almost exclusively in positively charged leptons.
To this end, it would be very useful for excluding new types of
physics if the full set of control samples of \cite{ATLASSUSY},
separated into subsamples with positively and negatively charged
leptons, would be made public.

\

Let us now turn  to the relevant studies of charge asymmetries.
To measure a charge asymmetry in a sample of events with one lepton is
straightforward.  Let $N_\pm$ to be the number of events in the sample
with an $\ell^\pm$, and let $N_{tot}=N_+ + N_-$ and $\Delta=N_+-N_-$.
Then the charge asymmetry is $A_C=\Delta/N_{tot}$.

In 2005, Bowen \cite{Bowen}, inspired by the forward-backward
asymmetry techniques used in single-top measurements at the Tevatron
\cite{BES, D0, CDF}, showed that charge asymmetries are useful in
extracting information about single-top production at the LHC.  He
noted that in a 14 TeV LHC event sample consisting of a lepton of
moderate $p_T$, moderate MET, and two or more jets, one of which is
$b$ tagged, the dominant contribution to the sample is from $t\bar t$,
which is nearly charge-symmetric.  At NLO $t\bar t$ production picks
up a small negative charge asymmetry (found in \cite{Ferrario:2008wm} to be no larger in magnitude than
$\sim 2\%$) in a subtle way: it arises from the intrinsic
forward-backward asymmetry in $q\bar q\to t\bar t$, which puts the
distribution of $\ell^+$ at higher $|\eta|$ than that of $\ell^-$.  A
small fraction of the $\ell^+$ events are then lost due to the
geometric acceptance of the detector.\footnote{Since CDF and D0 find
  that $A_{FB}^{t\bar t}$ is large, this small asymmetry may be
  enhanced; certainly this would be the case in the $W'$ model under
  consideration.  But because it arises from the subdominant $q\bar q$
  initial states, it remains small.  In addition it has a negative
  sign, opposite to our signal, so we are conservative in neglecting
  it here.  It can presumably be estimated, or bounded in absolute
  value, in data, using fully reconstructed $t\bar t$ events.}
Meanwhile, the largest contribution to a charge asymmetry in this
sample is from $t$-channel single top, with $W$-plus-jets
contributions coming a bit behind.  The reason $W$-plus-jets is so
small is that $b$-tagging is effective at rejecting it, combined with
the fact that events with charm jets actually have a negative
asymmetry that cancels off part of the positive asymmetry from the
other processes.

We first repeat this analysis at 7 TeV, accounting also for the new
contribution from the $W'$.  In the first numerical column of Table
\ref{tbl:effectofcuts} we show our estimates of cross-sections with
the event preselection cuts shown in Table \ref{tbl:preselectcuts};
note we also veto on a second isolated lepton.  (We will describe
the methods used for event simulation later.)  $W$-plus-jets (the
majority of which is $Wqg$) dominates the sample.

\begin{table}[h]
\begin{center}
\begin{tabular}[c]{|c||c|c|}
\hline
Item & $p_T$ & $|\eta|$
  \\  \hline
isolated $l^{\pm}$ & $\geq 20$ GeV & $\leq 2.1$ \\
MET (from $\nu$) & $\geq 20$ GeV & - \\
at least two  jets & $\geq 30$ GeV & $\leq 3.0$ \\ \hline
\end{tabular}

\caption{The preselection cuts for our samples. For
current LHC data sets there is no problem with triggering or
reconstruction at these values, but as we will see these cuts could be
raised if necessary.}
\label{tbl:preselectcuts}
\end{center}
\end{table}

In \cite{Bowen} the next and final stage was to apply a heavy-flavor
tag to at least one jet.  In this approach the key is to reduce
$W$-plus-jets as much as possible, and so one should apply a very
tight tag, with a very low mistag rate.  Let us get a feel for things
by first considering the effect of a heavy-flavor tag with a very
optimistic tagging rate.  (This would be appropriate for any attempt
to measure the SM single top contribution to the sample, since the
required statistics would be very large, by which point tagging would
be well-optimized.  It will not be appropriate for discussion of the
2010 data sample.)  The numbers in the second numerical column of
Table \ref{tbl:effectofcuts} reflect a rough estimate of the
cross-sections at the 7 TeV LHC that would result from a $70\%$
$b$-tag efficiency, a $15\%$ $c$-tag efficiency (conservatively low,
since $c$ quarks appear in the $W$-plus-jets background with a {\it
  negative} charge asymmetry), a $1\%$ efficiency for mistagging
light-quark jets, and a $3\%$ efficiency for $g$ jets (accounting both
for mistagging and for heavy-flavor tagging following $g\to c\bar c$
or $g\to b\bar b$ splitting.)  The reader may rescale the numbers in
Table \ref{tbl:effectofcuts} as desired.  At this stage the
$W$-plus-jet sample is as important as the SM single-top sample, and
the total asymmetry is small, just a few percent.  At the fiducial
point, the $W'$ model  would
dramatically increase the asymmetry, and dominate it even for
$F_S=0.25$.  Without the new signal, the SM asymmetry would be
about $4.5\%$.  In its presence, this would become $14.5\%$.  Given that
the sample has more than 2000 events, this is a signal of more than
4$\sigma$.

However, this is highly optimistic, especially in 2010.  First,
we have not even accounted correctly here for geometric acceptance;
tagging rates drop off to zero at $|\eta|=2.5$, and the rapidity
distribution of the signal's jets is quite wide.
More realistic heavy-flavor tagging and mistagging rates,
and proper treatment of their $p_T$ and $\eta$ dependence, would
reduce the significance.  Mistagging is likely to be
worse than we assumed here, especially in the presence of additional radiated jets,
and tagging efficiency is likely to be worse, especially for the $t$-channel
signal whose primary $b$ jet is often at quite high $p_T$.  And
the most serious problem could be the systematic error that comes
from a lack of knowledge of the tagging and mistagging rates at
high $p_T$.

\begin{table}[h]\begin{tabular}[c]{|c||c|c||c|c|c|c|}
\hline
Process & Preselection & Tag &  $H_T^{\ell\nu jj} > 350$
& $H_T^{\ell\nu jj} > 550$
  \\
 &  &  only &  GeV only&  GeV only
  \\ \hline
$W^+jj$ & 130 & 4.9 & 15   &  2.5   \\
$W^-jj$ & 71 & 2.6 & 6.5   &  1.1   \\ \hline
$W^+cj,W^+c\bar c$ & 18 & 2.7 & 1.5 &  0.11   \\
$W^-cj,W^-c\bar c$ & 24 & 3.6 & 2.2  &  0.41   \\ \hline
$W^+bb$ & 0.44 & 0.40 & 0.045   &  0.009   \\
$W^-bb$ & 0.26 & 0.24 & 0.017  &   0.003 
 \\ \hline \hline
SM NLO $t\bar b,tq,t\bar b q$& 3.5 & 2.5 &0.36   &  0.050   \\
SM NLO $\bar t b,\bar tq,\bar t b q$& 2.0 & 1.4 &0.13   &  0.014   \\ \hline \hline
SM NLO $t\bar t \to \ell^+$& 22 & 20 & 5.1   &   0.67  \\
SM NLO $t\bar t \to \ell^-$& 22 & 20 & 5.1   &   0.67  \\ \hline \hline \hline
New LO $td$           & 12 &  8.4    & 8.2   &  2.1    \\
New LO $\bar t\bar d$ & 0.90 &  0.63 & 0.61   &  0.15    \\ \hline
New LO $tb$           & 24 &  21    & 16.3   &  9.4  \\
New LO $\bar t\bar b$&  1.2 &  1.1 & 0.82   &  0.26   \\ \hline \hline
\end{tabular}
\caption{Cross-sections for SM backgrounds and $W'$-model signals in picobarns.
Results after preselection (see Table \ref{tbl:preselectcuts}),
after applying a heavy-flavor tag requirement along the lines of \cite{Bowen} (a
rough and optimistic estimate, with no $p_T$ or $\eta$ dependence),
and after applying cuts on $H_T^{\ell\nu jj}$ (with
no heavy-flavor tag) are shown. Details of the Monte Carlo simulation can be found in the main text.}
\label{tbl:effectofcuts}
\end{table}

Still,  the basic observation seems robust.
It seems likely that $F_S=0.25$, and perhaps
beyond, could be excluded through
the simple technique of \cite{Bowen}.

\

Because of the uncertainties surrounding the effectiveness of tagging,
we now consider an alternative and complementary approach, in which we
omit tagging and do a kinematic cut instead.  We will consider the
variable
\begin{equation}
H_T^{\ell\nu jj} = p_T^\ell + p_T^{j,1} + p_T^{j,2} + \MET
\end{equation}
where $p_T^{j,n}$ is the transverse momentum of the $n^{th}$-hardest jet,
$\MET$ is the missing transverse momentum in the event,
and the sum is a scalar sum of transverse momenta.
We will start by requiring $H_T^{\ell\nu jj}>$ 350 GeV (but {\it without}
applying a heavy-flavor tag).  This gives the numbers in the third numerical
column of Table \ref{tbl:effectofcuts}.

For this variable to be properly modeled, it is important that the first and second jet be
simulated correctly. In both signal and $t\bar t$, there are jets from
$t$ decays that have relatively low $p_T$, and ISR may easily give a
jet that is at higher $p_T$. In order to account for the additional
jets, we have generated a matched $t\bar t$ sample with up to one
additional jet using MadEvent \cite{MadEvent} with the implemented MLM
matching and the {\tt xqcut} variable set to 20 GeV. We then passed
the events through PYTHIA \cite{PYTHIA} for resonance decays
(including tops), showering and hadronization. Jets and geometric
acceptance were accounted for using PGS \cite{PGS} with the CMS
parameter set and $\Delta R=0.4$ cone jets. There are large error bars
associated with the use of this simulation tool, but we believe they
are no worse than other uncertainties that we are dealing with. The total $t \bar t$ cross section was normalized to the NLO result \cite{ttbarNLO} from MCFM \cite{ttbarNLOMCFM}.
The
signal was simulated using the {\tt usrmod} functionality in MadGraph
and run through PYTHIA and PGS in the same way.

For the $W$-plus-jets background we have been less careful, and have
performed only a parton level analysis, as the two leading jets are
simulated reasonably well in a $W$-plus-two-partons simulation.  We
included the effects of off-diagonal CKM matrix elements, as this has
a significant effect on $c$ quark production. We have used these LO
distributions to obtain the relative efficiencies of our kinematic
cuts on the $W$-plus-jets sample. This has known pitfalls, because
tails in distributions in variables such as $H_T^{\ell\nu jj}$ may be
larger after NLO corrections.  In a moment we will account for the unknown
normalization in the $W$-plus-jets contribution by rescaling it by a
constant that can be extracted from the data.  However, the NLO
effect on the charge asymmetry is not expected to be large, so we take
the LO result for the charge asymmetry after the kinematic cut as our best estimate.

  The
efficiency of SM single top under the $H_T^{\ell\nu jj}$ cut has also
been treated at LO parton-level, with the overall rates rescaled to
match the NLO cross section at 7 TeV \cite{TopNLOS,TopNLOT}.  Relative
to the large new signals, this process is too small to influence our
results.

Despite the large uncertainties in the $W$-plus-jets normalization,
the numbers in the last column of Table \ref{tbl:effectofcuts} already
show that the asymmetry in the SM and in the presence of the fiducial
signal are very different.  Even if we have underestimated the
$W$-plus-jets background by a factor of 4, the SM asymmetry is at
about 32\% with a statistical uncertainty of about 1.7\%, whereas in
the presence of the fiducial signal it is at 43\%, or 7$\sigma$ away
from the SM expectation.  We will see in a moment that
we have statistical sensitivity down to and potentially below $F_S=0.25$.

Systematic errors other than the overall normalization
of the $W$-plus-jets contribution may be very important.
These may arise from many sources,
including the top quark cross-section (which depends on the top quark
mass and also has NNLO corrections), the uncertainty in the
$W$-plus-jets asymmetry at NLO (which is believed to be small --- see
for example \cite{Stirling}), and the small top-quark charge asymmetry
discussed earlier.  There are also uncertainties in the signal, as we
have not used an NLO cross-section.  However, to the extent our
preselection efficiencies and that of the $H_T^{\ell\nu jj}$ cut do not
change too much at NLO, one can compensate for this effect by rescaling
the overall $W'$ coupling, which is directly absorbed into
$F_S$.

Certainly the largest uncertainty comes from normalizing the
$W$-plus-jets background subject to our simulation method and cuts.
We do not trust the normalization of our $W$-plus-jets sample, and
suspect it is significantly underestimated. Therefore we will multiply
the $W$-plus-jets background by a fudge factor $F_W$, which we will
imagine extracting from the data.  We may then consider the observed asymmetry,
and the observed cross-section of our sample $\sigma_{tot}$ after our
cuts, as a function of the two most important unknowns $F_W$ and $F_S$.  The observed
cross-section of the sample is quite sensitive to $F_W$.  By measuring
both $\sigma_{tot}$ and $A_C$, we can disambiguate, to a large extent,
the effect of $F_W$ and that of $F_S$.

\begin{figure}[h]
\centering
\includegraphics[width=90mm,height=90mm]{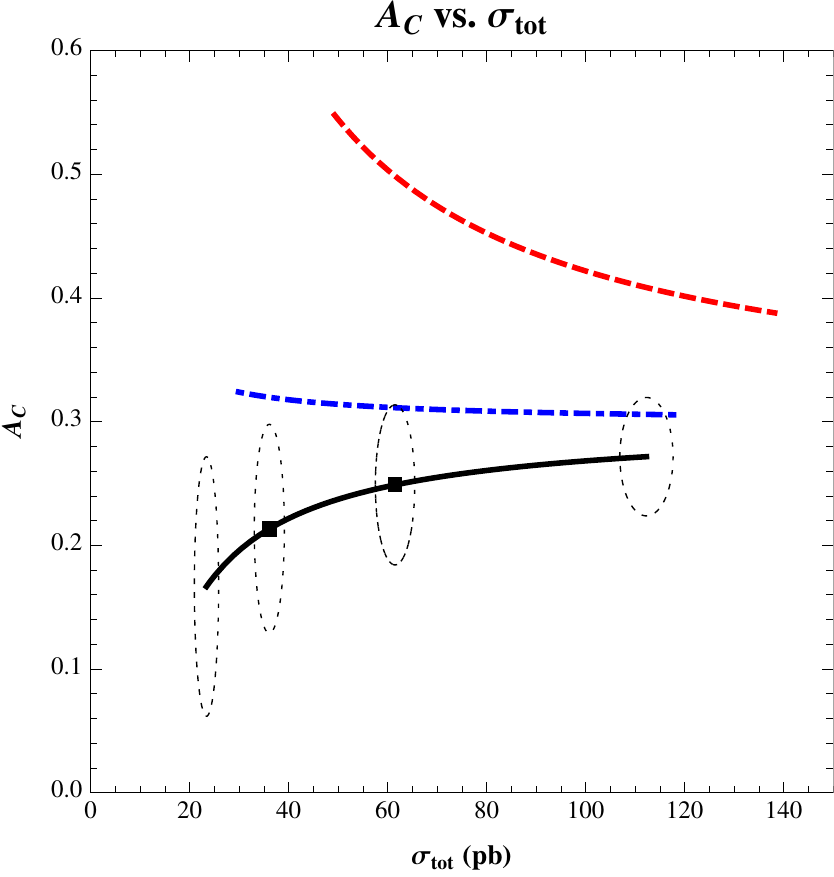}
\caption{We plot $A_C$ vs $\sigma_{tot}$ in three cases: the SM
  ($F_S=0$, lower solid curve), the SM plus 1/4 the fiducial signal
  ($F_S=0.25$, middle dot-dashed curve) and SM plus the fiducial
  signal ($F_S=1$, top dashed curve.)  (By the fiducial signal we mean
  the signal at $g_R=1.5$ and $M_{W'}=600$ GeV.) Curves run from
  $F_W=0.5$ at left to $F_W=4$ at right, where $F_W$ is the fudge
  factor for the $W$-plus-jets normalization.  Ellipses showing an
  estimate of 3$\sigma$ statistical uncertainties are shown for the SM
  and $F_W=0.5,1,2,4$.  Strong statistical separation is seen even for
  $F_S=0.25$, unless $F_W$ is very large.}
\label{fig:ACvsNtot}
\end{figure}

In Figure \ref{fig:ACvsNtot} we show $A_C$ versus the total cross
section $\sigma_{tot}=N_{tot}/(35 \ {\rm pb}^{-1})$ for the SM (solid
line, $F_S=0$), plotted from $F_W=0.5$ (at left) to $F_W=4$ (at
right).  We have also done so for the SM plus the fiducial signal
($F_S=1$), the top (dashed) curve, and for a reasonable target limit
of $F_S=0.25$, the middle (dash-dotted) curve.  For the SM and
$F_W=0.5,1,2,4$, we also show {\it three-sigma} statistical
error-ellipses corresponding to the statistical errors in $A_C$ and in
$\sigma_{tot}\times$ 35 pb$^{-1}$, ignoring correlations as well as
non-linearities in the $A_C$ uncertainties.  Clearly there is
excellent statistical separation everywhere except where $F_S$
approaches 0.25 and $F_W$ approaches 4.  Reaching this level of
sensitivity requires reducing the other systematic errors.  The
ongoing measurements of the $t\bar t$ cross-section will help pin down
the normalization of $t\bar t$ needed here.  Other measurements, such
as the cross-section and asymmetry in our preselection sample, for
which (at $F_S=0.25$) our signal makes no significant contribution,
can help determine the $W$-plus-jets cross-section given our $H_T$
cut.  In particular, it may be important to provide a bound from above
on $F_W$, using other measurements and theory.

\

So far we have taken an approach that tries to maximizes the size of the sample
and minimizes our errors in understanding tails of distributions.
Does it make sense to be more aggressive with the $H_T^{\ell\nu jj}$
cut?  We will see that we get only slightly better statistical sensitivity,
and there is a greater risk of systematic errors in the efficiency of
the cut.  But there may still be benefits.

\begin{figure}[h]
\centering
\includegraphics[width=90mm,height=90mm]{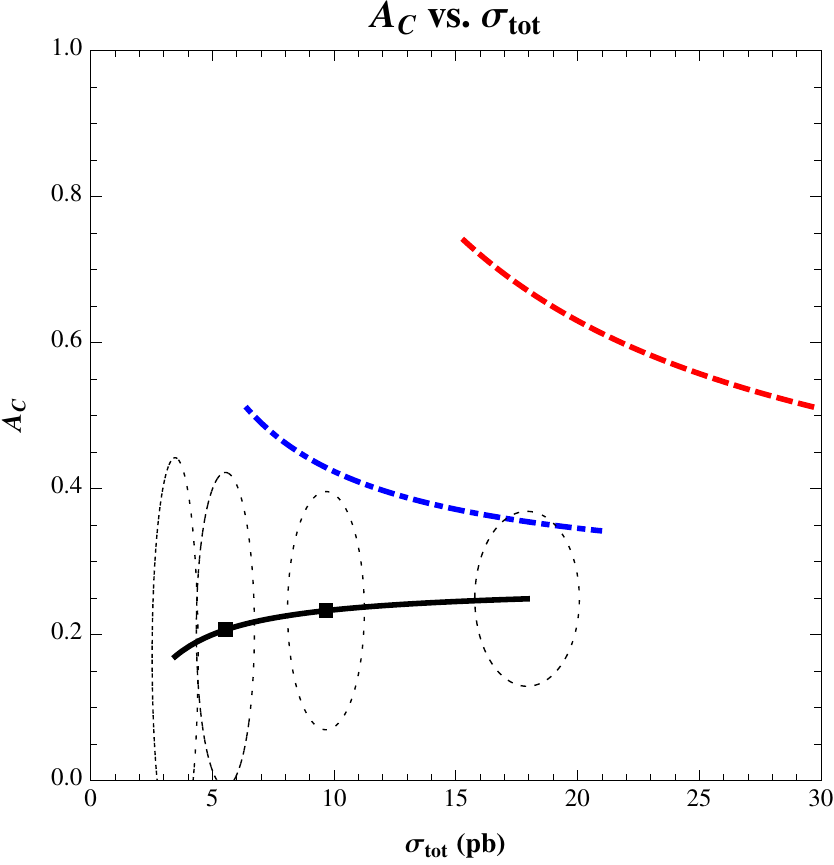}
\caption{As in Figure \ref{fig:ACvsNtot},
but with $H_T^{\ell\nu jj}>550$ GeV, and plotted from $F_W=0.5$ at left
to $F_W=4$.
Ellipses showing an
  estimate of 3$\sigma$ statistical uncertainties are shown for the SM
  and $F_W=0.5,1,2,4$.}
\label{fig:ACvsStot550}
\end{figure}

In the final column of Table \ref{tbl:effectofcuts} we repeat the
previous exercise while requiring $H_T^{\ell\nu jj}>550$ GeV.  Note
that the composition of the sample has significantly changed.  The
$t\bar t$ fraction is reduced, due presumably to the fall in the $gg$
parton luminosity.  The corresponding plot of $A_C$ versus
$\sigma_{tot}$ is given in Figure \ref{fig:ACvsStot550}.  Again we
allow $F_W$ to vary up to 4; note that the appropriate value of $F_W$
for this figure will not be equal to that for the previous figure, as
the error in our estimate of $W$-plus-jets will vary with the
kinematic cuts.  We see for $F_S=0.25$, statistical power improves
for small $F_W$, though not for $F_W\to 4$.

Since even $t\bar t$ is well out on its high-momentum tail, we should
worry about how uncertain is the efficiency of our kinematic cut.
Though we use a $t\bar t(j)$ matched sample passed through PYTHIA to
help us model that tail, still one must not take our numbers for the
$t\bar t$ background too seriously.  But here heavy-flavor tagging
becomes useful.

The first point is that the SM in this range produces a charge asymmetry
that comes dominantly from a contribution that is $b$-poor.  Meanwhile
the SM plus the $F_S=0.25$ signal produces a slightly larger asymmetry
due to a $b$-rich contribution.  Therefore, when a heavy-flavor tag is
applied, the asymmetry will generally {\it decrease} significantly if
$F_S=0$ (pure SM), and {\it increase} if $F_S=0.25$.  Moreover,
especially at large $F_W$ where separation of signal and background is
worst before tagging, the reduction in $\sigma_{tot}$ after tagging is
much greater in the SM than in the presence of an $F_S=0.25$ signal.
Neither of these statements is true for the $H_T^{\ell\nu jj}>350 $ GeV
sample, because there is too much $t\bar t$ left after tagging, which
dilutes the asymmetry of the signal and contributes significantly to
$\sigma_{tot}$.

The asymmetry and cross-section after a single tag is required are
somewhat sensitive to the amount of $t\bar t$ remaining in the sample.
But the $t\bar t$ fraction can be estimated (or at least bounded from above)
by also considering the sample with two tags.  Double-tagging will
completely remove $W$-plus-jets and leave a combination of $t\bar t$
and some remaining $t$-channel signal.  The asymmetry and
cross-section can again be measured, constraining  the $t\bar t$ fraction
independently.

Whether these methods allow any improvement in the statistical
significance of the measurement is very sensitive to the details of
the heavy-flavor-tagging technique.  Many such techniques could be
imagined.  For instance, for the single-tag sample, one might tag only
the two hardest jets, which would reduce $t\bar t$ and avoid overly
large mistag rates in $W$-plus-jets.  Its efficiency for signal would
need study.  Alternatively, one might only apply tagging to the
second-highest-$p_T$ jet, since this jet is often a $b$ jet in signal,
and has low enough $p_T$ to be in the ``sweet spot'' for tagging with
high efficiency.  In contrast, the highest-$p_T$ jet, also often a
$b$, is at such high $p_T$ (typically $>250$ GeV) that its tagging
efficiency is not that high.  Meanwhile the second jet in the
$W$-plus-jets background, unlike the hardest jet, is also often at low
enough $p_T$ that mistagging rates may be near their low point.  There
will be some loss of signal and increased theoretical errors compared
to a technique that tags more widely, but the corresponding reduction
in mistagging of the background, and in the uncertainties in tagging
efficiencies, may be worth it.

Just to give a feel for the numbers, let us consider an example.
Suppose mistagging of $W$-plus-jets could be brought down to 3\% per
event (10\% for events with charm and 60\% for $Wbb$), and if tagging
of $t\bar t$ events were of order 60\%, with signal events tagged at
40\% probability.  Now suppose that a charge asymmetry of 30\%, in a sample
with a cross-section of 18 pb is measured.  This is statistically
consistent with $F_S=0, F_W=4$ or $F_S=0.25,F_W=3.3$, whose
asymmetries have central values of 25\% and 35\% respectively.  Then
for the SM alone, after tagging, we expect a cross-section of 1.7 pb
and an asymmetry of 14\%.  In the presence of a $F_S=0.25$ signal, the
asymmetry will instead move up to 41\%, with a cross-section of 2.7
pb.  With 35 inverse pb, the total number of events is of order 50 --
100, so statistical uncertainties are large.  But progress has still
been made.  The progress is easily lost, however, if mistagging is a
more serious problem, or if tagging of the signal is significantly
worse.

\

Now let us put these results together.  We have seen that by combining
tagging and kinematic cuts we can get at least 2.5 $\sigma$
statistical sensitivity to $F_S=0.25$, in several different ways.
While these different ways are not independent, they do have very
different combinations of backgrounds, and different sources of
systematic errors.  Properly combined, they should allow for even
better sensitivity.

Surely the best way to do this, including all the information, is to
simultaneously study the differential distribution versus
$H_T^{\ell\nu jj}$ of {\it both} $A_C$ {\it and} the total number of
events, both before and after the application of a wisely-chosen
heavy-flavor-tagging method.  It should be possible to discover or
exclude the model even well below $F_S=0.25$. This is an important
range to aim at, as we have emphasized.

This said, we should add one caveat.  We have shown that the $W'$
model of \cite{SZ} can easily be discovered or excluded down well
below its fiducial cross-sections.  But the absence of a signal might
merely imply that the $W'ub$ coupling is absent from the model.  While
the model then could not explain the anomalies in the $B$ system, part
of its original motivation, it might still explain the anomalous
$A_{FB}^{t\bar t}$ in top pair production through a $W'td$ coupling.
In this case, distortions in $t\bar tj$ samples due to $t W'$
production may be the dominant observable signal in such a model.
Discovering such a model will be somewhat more challenging, but would
still not take long, given the large coupling of the $W'$ and the
kinematic structures associated with its large mass.

Interestingly, charge asymmetries will have a crucial role to play in
this case as well.  Although there are equal numbers of positively and
negatively charged leptons produced in $tW'$ events, they will have
very different $p_T$ distributions.  This is because the cross-section
for $tW^{'-}$ is very much larger than that for $\bar t W^{'+}$, and
the $t$ from the $W'$ decay will have much higher $p_T$ than the $t$
produced directly.  Therefore a plot of the lepton $p_T$ will be very
different for the two lepton charges.  If a sufficiently clean sample
can be obtained, for example by requiring two $b$ tags, the
backgrounds, mostly $t\bar t$, will show a much smaller difference.

Similarly, as we mentioned in our introduction,
there has been interest recently in models
with a $Z'$ that couples to $ut$, and allow for the highly asymmetric
process $uu\to tt$. In this case same-sign leptons that are mainly
of positive charge rather than negative are a clear sign of a process
from a $uu$ initial state; the plus-two charge of the $pp$ collision
is entirely transferred to leptons.  There will also of course be
an asymmetry in one-lepton samples, though the backgrounds to same-sign
dilepton events are much smaller.\footnote{After this paper was completed,
it was pointed out \cite{Tait} that the one-lepton asymmetry would actually
be quite sensitive, due to its larger statistics.}

\

In this paper, we have considered a $W'$ coupling both to right-handed
$ub$ and $td$, which has been suggested \cite{SZ} as a solution to
both the $A_{FB}^{t\bar t}$ excess and various puzzles in $B$ mesons.
We noted that the model has a large cross-section for single-top
production.  At fiducial coupling and mass of $g_R=1.5$ and
$M_{W'}=600$ GeV, the model is probably ruled out by Tevatron
single-top measurements \cite{GroupST}, although a quantitative
assessment requires use of the multivariate techniques employed by the
detector collaborations.  It also appears the fiducial parameter
region is excluded by existing LHC measurements, as in the control
regions of \cite{ATLASSUSY}.  However, the model with somewhat smaller
$g_R$ and/or $1/M_{W'}$ is harder to exclude with existing public
results, and could still serve to explain the observed large
$A_{FB}^{t\bar t}$ if this anomaly turns out to be currently
overestimated.  We have argued (inspired by \cite{Bowen}) that even
with a rate reduced by four or more relative to the fiducial model's
LO rate, single-top-quark production in this model creates an
significant excess charge asymmetry in a transparent sample with
simple kinematic cuts and/or heavy-flavor tagging.  Our conclusion is
that the current 2010 data sets at ATLAS and CMS of $\sim 35$ inverse
picobarns apparently suffice to detect even this reduced signal, or to
strongly disfavor the single-top process down to levels very
significantly smaller than predicted by the fiducial model.

We would also like to emphasize the model-independent value of
this measurement.  We hope that any analysis along these lines
is presented in a model-independent fashion, as well as in the form of
limits on the specific $W'$ model of \cite{SZ}.

It should be clear that the method we have outlined, and ones of a
similar form, will work on any large charge-asymmetric signals
that produce leptons, neutrinos and jets, and perhaps
$b$ jets.  Our particular set of strategies will continue to be effective at
higher luminosity and with higher kinematic cuts.  We would
argue that these techniques should be in the standard toolkit of the
LHC experimental community: that at every significant step in
increased integrated luminosity, it is important to produce a
simultaneous analysis of differential charge asymmetries and
cross-sections versus $m_{{\rm eff}}$, $H_T$, or other kinematic
variables, for different numbers of heavy-flavor-tagged jets.  These
analyses will be significantly more powerful than analysis of differential
cross-sections alone.  Here we echo previous general arguments to this
effect \cite{Pheno2005, Stirling}.  The methods that we have proposed,
and others along similar lines, will continue to be useful throughout
the lifetime of the LHC. \\

{\bf Note Added} --- After this article was completed, results of powerful
searches for Standard Model single-top production were announced by
both CMS and ATLAS \cite{CMSt,ATLASt}.  We have reconsidered the
situation in light of these new analyses.

As we suggested would be possible, limits from the LHC on the $W'$
model of \cite{SZ} appear now to reach values of order $F_S=0.25$.  We
conclude this not from the quoted limits in \cite{CMSt,ATLASt} on the
SM single-top cross-section, which were obtained by optimizing for that
rather idiosyncratic process, but from plots characterizing the
preselection samples.  The most useful plots are Figures 3c and 3h of
\cite{ATLASt} and Figure 15 of \cite{CMSt}.  By roughly reproducing
these figures, and considering the size and shape of the signal from
the $W'$ model, we estimate that with $F_S=0.25$ the signal would be
detected on the tails of these distributions, even with pessimistic
efficiencies for lepton identification and heavy-flavor tagging.

But any accurate estimate of excluded values of $F_S$ would require
additional information about heavy-flavor tagging rates.  As
emphasized above, the precise limits depend sensitively on the tagging
efficiency and mistagging rate for high $p_T$ jets, and on how well
these are known.  This information was not provided in the ATLAS and
CMS papers (only average information on the working point was given,
and this is appropriate at lower values of $p_T$ than is relevant for
the high $H_T$ or $\hat s$ region.)  We would encourage both ATLAS and
CMS to provide more detailed information about tagging methods in
future publications, so that the reported results can be more widely
used.

One important fact we learn from the small numbers of events with two
jets, at high $H_T$ in the ATLAS figure\footnote{Note there is one
  event at $H_T>400$ GeV not shown in Figure 3h of \cite{ATLASt}; this
  can be inferred from the corresponding table in the text.  We thank
  Kyle Cranmer of ATLAS for helpful discussions.}  and at high $\hat
s$ in the CMS figure, is that the $W$-plus-jets background is small
after tagging.  However, this could have two possible causes.  It
could be that the $Wjj$ cross-section (and therefore $F_W$, in our
notation) is not much larger than given by our leading order estimate.
If this is the case, then, as our figures suggest, these LHC studies
will have strong sensitivity to the $W'$ model.  But if instead the
small background is due to tight tagging, with a low mistag rate but a
correspondingly relatively low $b$-tagging rate, then considerable
sensitivity may have been lost.

Lacking the information, we will not try to explore these searches
further at this time.  Instead, we return to the analysis that we
suggested above, and compare general aspects of the different
strategies.  We believe that if a method closer to the one we have
proposed were adopted, it would allow for even stronger limits on the
$W'$ model, and perhaps on many other phenomena.

A key difference is that we are not seeking a region of zero
background, because, in using the asymmetry as well as the
cross-section, we do not need it.  In particular, it appears
disadvantageous to consider only the two-jet sample, as was 
deemed necessary
for the SM single-top searches in \cite{CMSt,ATLASt}.

Moreover, we would suggest comparing the samples with the tighter and
looser kinematic cuts, before and after tagging, to get even more
sensitivity.  Essentially, in our language, this pins down $F_W$, the
$W$ contribution to the sample, thus determining the expected SM
cross-section and asymmetry to a greater degree.

To demonstrate this, we present two new figures, which are similar to
our original ones but with the following differences.

First, tagging is imposed.  We choose a mistag rate of $5\%$ per $Wjj$
event ({\it not} per jet), $15\%$ on average per $Wcj$ and $Wcc$
event, $75\%$ for $Wbb$, $t\bar t$ and SM single top, and $50\%$ for
our signal (recalling that much of it has two $b$ jets but often one
jet has $p_T>200$ GeV, where tagging is degraded.)  This may or may not
be optimistic, but we note that $Wjj$ is sufficiently small, after
requiring $H_T^{\ell\nu jj}>550$ GeV, that moderate changes in
mistagging do not drastically change the result.  Consequently one can
roughly adjust these figures for changes in tagging by rescaling the
signal (background) almost linearly (quadratically) with the
$b$-tagging efficiency.

Second, $F_S=1$ is thoroughly excluded, so we now show $F_S=0.25$
and $F_S=0.1$.

Third, the statistics is so low, after tagging, for the background-only
case that we can better illustrate the uncertainties by showing
1$\sigma$ and 3$\sigma$ statistical fluctuations on the
signal-plus-background hypothesis.  We center the contours on the
point $F_W=1$, $F_S=0.1$.

\begin{figure}[h]
\centering
\includegraphics[width=90mm,height=90mm]{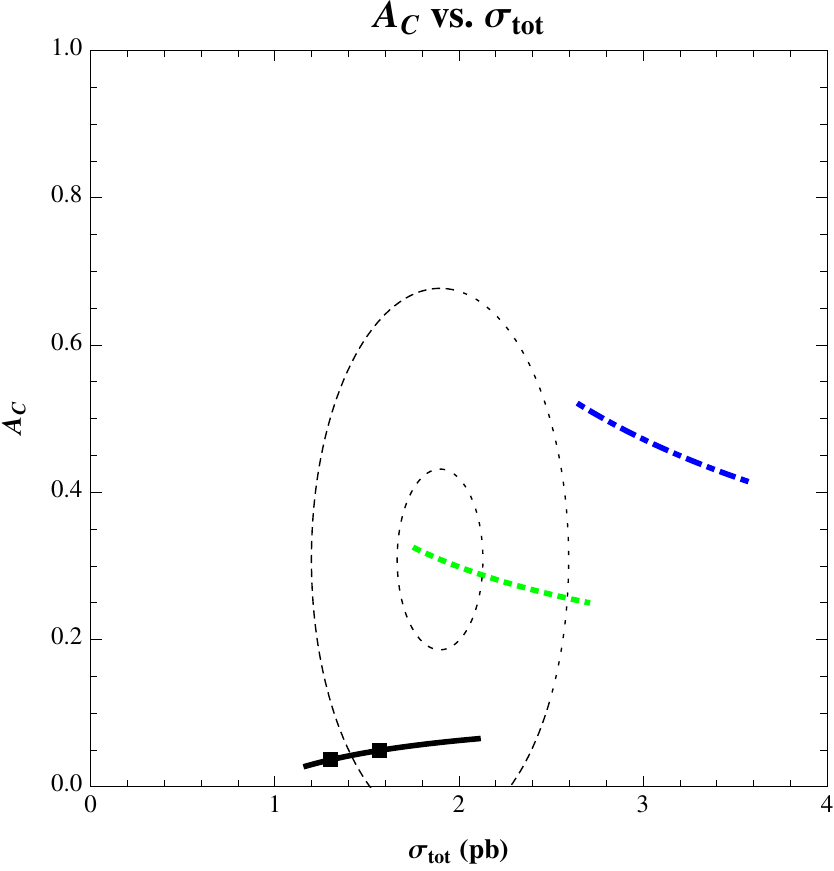}
\caption{We plot $A_C$ vs $\sigma_{tot}$, after requiring
$H_T^{\ell\nu jj}>$ 550 GeV and imposing a heavy-flavor tag (see text for details),
in three cases: the SM
  ($F_S=0$, lower solid curve), the SM plus 1/10 the fiducial signal
  ($F_S=0.1$, middle dotted curve) and the SM plus 1/4 the fiducial signal
  ($F_S=0.25$, top dot-dashed curve.)  Notice that we have not plotted
the same quantities as in Figs. \ref{fig:ACvsNtot} and \ref{fig:ACvsStot550}.   Curves run from
  $F_W=0.5$ at left to $F_W=4$ at right, with dots on the SM curve at
  $F_W=1$ and $2$, where $F_W$ is the fudge
  factor for the $W$-plus-jets normalization.  Ellipses showing an
  estimate of 1$\sigma$ and 3$\sigma$ statistical uncertainties are shown for the
case of $F_S=0.1$ and $F_W=1$.}
\label{fig:ACvsStot550tag}
\end{figure}

\begin{figure}[h]
\centering
\includegraphics[width=90mm,height=90mm]{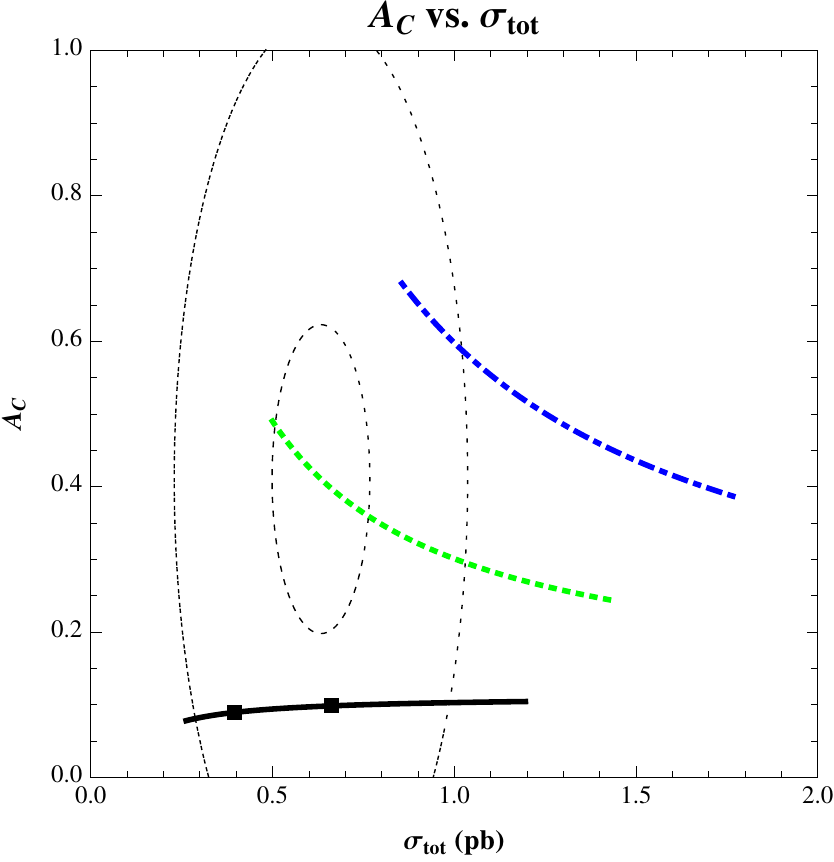}
\caption{As in Fig.~\ref{fig:ACvsStot550tag}, but having also required two and
only two jets (see text for details.)  Notice the background is
lower, but sensitivity is lost in our two-dimensional analysis.}
\label{fig:ACvsStot550tag2jet}
\end{figure}

Finally, the two figures illustrate the difference between taking a
two-or-more jet sample, shown in Fig.~\ref{fig:ACvsStot550tag}, and
requiring two and only two jets, shown in
Fig.~\ref{fig:ACvsStot550tag2jet}.  The reduction in $t\bar t$
background arising from the two-jet restriction improves a pure
counting experiment, but the remaining statistics is too low for $A_C$
to be a useful variable.  If instead one aims at dividing the events
and comparing positively and negatively charged lepton samples, the
two-jet restriction degrades sensitivity.  (Similarly, very tight
tagging requirements may well be counter-productive.)  Our figures
suggest that sensitivity would be improved --- especially if one
measures $F_W$ using another method, such as determining it from the
untagged sample --- with the looser event requirements.

As an aside, we note that Figure 3 of \cite{CMSt} and Figure 15
of \cite{ATLASt} combine positively and negatively charged leptons,
rather than showing them separately.  For the future, we encourage the
CMS and ATLAS collaborations to present plots separated by charge,
both in searches such as this one where
charge asymmetries might obviously be of interest, and in other cases
where the absence of a large charge asymmetry may be powerful in
excluding various types of new physics.

We conclude this Note-Added with two messages.  First, we strongly
encourage the CMS and ATLAS collaborations to revisit their single-top
analyses\footnote{For example, it should be possible to reanalyze them
  using the RECAST method \cite{RECAST}.}  to put proper limits on
this model, with its two parameters $g_R$ and $M_{W'}$.  We expect that
the model will be ruled out throughout the region in which it
would have served its original purposes.  Second, we urge the
collaborations to consider the value of using more inclusive samples, as we
have suggested in our article, compensating for the increase in background
with the power of the charge asymmetry.

\

The authors thank Y.~Gershtein, E.~Halkiadakis, J.~Ruderman, D.~Shih,
N.~Toro and G.~Watts for comments and conversations.  We also thank K.~Cranmer
and D.~Tardif for discussions relevant to the Note Added.
The work of C.K. and M.J.S. was supported by NSF grant PHY-0904069 and
by DOE grant DE-FG02-96ER40959. The work of N.C. is supported by NSF
grant PHY-0907744.


\begin{thebibliography}{99}

\bibitem{CDFafb}
  T.~Aaltonen {\it et al.}  [CDF Collaboration],
   ``Forward-Backward Asymmetry in Top Quark Production in  $p\bar{p}$
  Collisions at $\sqrt{s}$ =1.96 TeV,''
  Phys.\ Rev.\ Lett.\  {\bf 101}, 202001 (2008)
  [arXiv:0806.2472 [hep-ex]]


\bibitem{D0afb}
  V.~M.~Abazov {\it et al.}  [D0 Collaboration],
   ``First measurement of the forward-backward charge asymmetry in top quark
  pair production,''
  Phys.\ Rev.\ Lett.\  {\bf 100}, 142002 (2008)
  [arXiv:0712.0851 [hep-ex]].

\bibitem{CDFnew}
  T.~Aaltonen {\it et al.}  [The CDF Collaboration],
  ``Evidence for a Mass Dependent Forward-Backward Asymmetry in Top Quark Pair
  Production,''
  arXiv:1101.0034 [hep-ex].

\bibitem{Hitoshi}
  S.~Jung, H.~Murayama, A.~Pierce, J.~D.~Wells,
  Phys.\ Rev.\  {\bf D81}, 015004 (2010).
  [arXiv:0907.4112 [hep-ph]].



\bibitem{Wagner}
  Q.~-H.~Cao, D.~McKeen, J.~L.~Rosner, G.~Shaughnessy, C.~E.~M.~Wagner,
  Phys.\ Rev.\  {\bf D81}, 114004 (2010).
  [arXiv:1003.3461 [hep-ph]].


\bibitem{SZ}
  J.~Shelton, K.~M.~Zurek,
  ``A Theory for Maximal Flavor Violation,''
  [arXiv:1101.5392 [hep-ph]].

  \bibitem{Cheung:2009ch}
  K.~Cheung, W.~Y.~Keung and T.~C.~Yuan,
  Phys.\ Lett.\  B {\bf 682}, 287 (2009)
  [arXiv:0908.2589 [hep-ph]].

\bibitem{Cheung}
  K.~Cheung and T.~C.~Yuan,
 ``Top Quark Forward-Backward Asymmetry in the Large Invariant Mass Region,''
  arXiv:1101.1445 [hep-ph].


\bibitem{Barger:2010mw}
  V.~Barger, W.~Y.~Keung and C.~T.~Yu,
  Phys.\ Rev.\  D {\bf 81}, 113009 (2010)
  [arXiv:1002.1048 [hep-ph]].




  \bibitem{D0dimuon}
  V.~M.~Abazov {\it et al.}  [D0 Collaboration],
  ``Evidence for an anomalous like-sign dimuon charge asymmetry,''
  Phys.\ Rev.\  D {\bf 82}, 032001 (2010)
  [arXiv:1005.2757 [hep-ex]].

  \bibitem{D0Bs}
  V.~M.~Abazov {\it et al.}  [D0 Collaboration],
  ``Measurement of $B^0_{s}$ mixing parameters from the flavor-tagged decay
  $B^0_{s} \to J/\psi \phi$,''
  Phys.\ Rev.\ Lett.\  {\bf 101}, 241801 (2008)
  [arXiv:0802.2255 [hep-ex]].

  \bibitem{CDFBs}
  L.~Oakes  [CDF Collaboration],
  ``Measurement of $\beta_s$ at CDF,''
  arXiv:1102.0436 [hep-ex].

  \bibitem{Lunghi}
  E.~Lunghi, A.~Soni,
  ``Possible evidence for the breakdown of the CKM-paradigm of CP-violation,''
  Phys.\ Lett.\  {\bf B697}, 323-328 (2011).
  [arXiv:1010.6069 [hep-ph]].

\bibitem{Ligeti}
  Z.~Ligeti, M.~Papucci, G.~Perez, J.~Zupan,
  ``Implication s of the dimuon CP asymmetry in $B_{d,s}$ decays,''
  Phys.\ Rev.\ Lett.\  {\bf 105}, 131601 (2010).
  [arXiv:1006.0432 [hep-ph]].

  \bibitem{Barger}
  V.~Barger, L.~Everett, J.~Jiang, P.~Langacker, T.~Liu and C.~Wagner,
  ``Family Non-universal $U(1)^\prime$ Gauge Symmetries and $b\to s$
  Transitions,''
  Phys.\ Rev.\  D {\bf 80}, 055008 (2009)
  [arXiv:0902.4507 [hep-ph]].

\bibitem{Bowen}
  M.~T.~Bowen,
  ``Using charge asymmetries to measure single top quark production at the LHC,''
  Phys.\ Rev.\  {\bf D73}, 097501 (2006).
  [hep-ph/0503110].

\bibitem{BES}
  M.~T.~Bowen, S.~D.~Ellis, M.~J.~Strassler,
  ``In search of lonely top quarks at the Tevatron,''
  Phys.\ Rev.\  {\bf D72}, 074016 (2005).
  [hep-ph/0412223].


\bibitem{asymsusy}
 C. Albajar et. al. in Aachen LHC
Collider Workshop, CERN90-10 (1990); F. Pauss, ibid.\\
  H.~Baer, X.~Tata and J.~Woodside,
  ``Multi - lepton signals from supersymmetry at hadron super colliders,''
 Phys.\ Rev.\  D {\bf 45}, 142 (1992). \\
  H.~Baer, M.~Bisset, X.~Tata and J.~Woodside,
  ``Supercollider signals from gluino and squark decays to Higgs bosons,''
  Phys.\ Rev.\  D {\bf 46}, 303 (1992). \\
    H.~Baer, C.~h.~Chen, F.~Paige and X.~Tata,
  ``Signals for Minimal Supergravity at the CERN Large Hadron Collider II:
 Multilepton Channels,''
    Phys.\ Rev.\  D {\bf 53}, 6241 (1996)
  [arXiv:hep-ph/9512383]. \\
S.~Muanza,
``Using Charge Asymmetry in the Search for Chargino-Neutralino Pairs at the LHC,''
GDR-S-076 (2000). Available online at
\newline
{\tt
susy.in2p3.fr/GDR-Notes/GDR\_SUSY\_PUBLIC/GDR-S-076.ps} \\
  A.~J.~Barr,
  ``Determining the spin of supersymmetric particles at the LHC using lepton
  charge asymmetry,''
  Phys.\ Lett.\ B {\bf 596}, 205 (2004)
  [arXiv:hep-ph/0405052]. \\
  T.~Goto, K.~Kawagoe and M.~M.~Nojiri,
  ``Study of the slepton non-universality at the CERN Large Hadron Collider,''
  Phys.\ Rev.\ D {\bf 70}, 075016 (2004)
  [Erratum-ibid.\ D {\bf 71}, 059902 (2005)]
  [arXiv:hep-ph/0406317]. \\
  T.~Goto,
  ``Neutralino polarization effect in the squark cascade decay at LHC,''
  arXiv:hep-ph/0411360.

\bibitem{Slabospitsky}
  G.~V.~Jikia, S.~R.~Slabospitsky,
  ``Single top production at hadron UNK collider,''
  Sov.\ J.\ Nucl.\ Phys.\  {\bf 55}, 1387-1392 (1992).


\bibitem{Pheno2005}
M.~J.~Strassler,
``New/Old Methods at LHC for Single Top and Beyond,''
talk presented at Pheno2005, Madison, WI (2005). Available online at
\newline
{\small{\tt http://www.physics.rutgers.edu/$\sim$strassler/conference\_talks/ACPheno.pdf}}



\bibitem{Stirling}
  C.~H.~Kom and W.~J.~Stirling,
  ``Charge asymmetry in W + jets production at the LHC,''
  Eur.\ Phys.\ J.\  C {\bf 69}, 67 (2010)
  [arXiv:1004.3404 [hep-ph]].


 \bibitem{GroupST}
  T.~E.~W.~Group  [CDF and D0 Collaboration],
  ``Combination of CDF and D0 Measurements of the Single Top Production Cross
  Section,''
  arXiv:0908.2171 [hep-ex].

  \bibitem{ATLASSUSY}
J.~B.~G.~da Costa {\it et al.}  [Atlas Collaboration],
  ``Search for supersymmetry using final states with one lepton, jets, and
  missing transverse momentum with the ATLAS detector in sqrt{s} = 7 TeV pp,''
  arXiv:1102.2357 [hep-ex].

\bibitem{ATLASTalk}
A.~Farbin [Atlas Collaboration],
``Recent SUSY Searches by ATLAS,''
talk presented at Aspen Workshop on New Data from the Energy Frontier (2011).



 \bibitem{D0}
  V.~M.~Abazov {\it et al.}  [D0 Collaboration],
  ``Evidence for production of single top quarks and first direct  measurement
  of |V(tb)|,''
  Phys.\ Rev.\ Lett.\  {\bf 98}, 181802 (2007)
  [arXiv:hep-ex/0612052].

  \bibitem{CDF}
  T.~Aaltonen {\it et al.}  [CDF Collaboration],
  ``Measurement of the Single Top Quark Production Cross Section at CDF,''
  Phys.\ Rev.\ Lett.\  {\bf 101}, 252001 (2008)
  [arXiv:0809.2581 [hep-ex]].


\bibitem{Ferrario:2008wm}
  P.~Ferrario and G.~Rodrigo,
  Phys.\ Rev.\  D {\bf 78}, 094018 (2008)
  [arXiv:0809.3354 [hep-ph]].

\bibitem{MadEvent}
  F.~Maltoni and T.~Stelzer,
  ``MadEvent: Automatic event generation with MadGraph,''
  JHEP {\bf 0302}, 027 (2003)
  [arXiv:hep-ph/0208156];
  J.~Alwall {\it et al.},
  ``MadGraph/MadEvent v4: The New Web Generation,''
  JHEP {\bf 0709}, 028 (2007)
  [arXiv:0706.2334 [hep-ph]].

 \bibitem{PYTHIA}
  T.~Sjostrand, S.~Mrenna and P.~Z.~Skands,
  ``PYTHIA 6.4 Physics and Manual,''
  JHEP {\bf 0605}, 026 (2006)
  [arXiv:hep-ph/0603175].

 \bibitem{PGS}
  J.~Conway {\em et~al.},
  ``{PGS 4: Pretty Good Simulation of high energy collisions},'' 2006,
\newline
{\small  {\tt www.physics.ucdavis.edu/$\sim$conway/research/software/pgs/pgs4-general.htm}}



\bibitem{ttbarNLO}
  V.~Khachatryan {\it et al.}  [CMS Collaboration],
  ``First Measurement of the Cross Section for Top-Quark Pair Production in
  Proton-Proton Collisions at sqrt(s)=7 TeV,''
  Phys.\ Lett.\  B {\bf 695}, 424 (2011)
  [arXiv:1010.5994 [hep-ex]].

\bibitem{ttbarNLOMCFM}
  J.~M.~Campbell and R.~K.~Ellis,
  ``MCFM for the Tevatron and the LHC,''
  Nucl.\ Phys.\ Proc.\ Suppl.\  {\bf 205-206}, 10 (2010)
  [arXiv:1007.3492 [hep-ph]].



   \bibitem{TopNLOS}
  S.~Heim, Q.~H.~Cao, R.~Schwienhorst and C.~P.~Yuan,
  ``Next-to-leading order QCD corrections to s-channel single top quark
  production and decay at the LHC,''
  Phys.\ Rev.\  D {\bf 81}, 034005 (2010)
  [arXiv:0911.0620 [hep-ph]].


  \bibitem{TopNLOT}
  R.~Schwienhorst, C.~-P.~Yuan, C.~Mueller, Q.~-H.~Cao,
  ``Single top quark production and decay in the $t$-channel at next-to-leading order at the LHC,''
  Phys.\ Rev.\  {\bf D83}, 034019 (2011).
  [arXiv:1012.5132 [hep-ph]].

\bibitem{Tait}
  A.~Rajaraman, Z.~'e.~Surujon, T.~M.~P.~Tait,
  ``Asymmetric Leptons for Asymmetric Tops,''
  
  [arXiv:1104.0947 [hep-ph]].

\bibitem{CMSt}
The CMS collaboration,
``Measurement of the single-top t-channel cross section in pp collisions at sqrt(s)=7 TeV'', CMS-PAS-TOP-10-008 (2011)

\bibitem{ATLASt}
The ATLAS collaboration,
``Searches for Single Top-Quark Production with the ATLAS Detector in pp Collisions at sqrt(s) = 7TeV'', ATLAS-CONF-2011-027 (2011).



\bibitem{RECAST}
  K.~Cranmer, I.~Yavin,
  ``RECAST: Extending the Impact of Existing Analyses,''
  JHEP {\bf 1104}, 038 (2011)
  [arXiv:1010.2506 [hep-ex]].




\end{thebibliography}
\end{document}